# Nuclear Data and Fuel/Assembly Manufacturing Uncertainties Analysis and Preliminary Validation of SUACL

JiaYi-Xu[a], Xu Bo-Ma[a,*], Fan Lu[a] and Yi Xue-Chen[a]
[a]*School of Nuclear Science and Engineering, North China Electric Power University, Beijing 102206, China*
*Corresponding author:maxb917@163.com.Xu Bo-Ma

## 1. Introduction

As the sensitivity and uncertainty analysis of nuclear system can provide more confident bounds for the Best-estimate Prediction used to assess the performance and safety of nuclear plant, the uncertainty and sensitivity analysis has been a component of analysis of nuclear system.There are hundreds of uncertainty sources in reactor physics calculation, only epistemic or subjective uncertainties can be analyzed. And the imprecise input parameters are recognized as epistemicuncertainties [1].The objective of uncertainty analysis is to quantify the uncertainty of output derived from the relative inputs u n certainty[2].The cross section and parameters of fuel/assembly manufacturing are the fundamental inputs of the neutron transport equation.Theimprecise nuclear data has been treated as one of the major uncertainty sources [3]. Butthe parameters of fuel/assembly manufacturinguncertainties are paid little attention. Both the cross section uncertainty and the uncertainty of parameters of fuel/assembly manufacturing are analyzed in this paper.

Generally, all cross section can be divided into the basic and integralones.The integral cross sections are composed of basic cross sections. The cross section uncertainties are stored in their covariance data. Different evaluate approaches used in nuclear data library may lead to the different of analysis result.For verify the effect, the covariance library uses the covariance data from NJOY99 based onENDF/B-VII.1 and JENDL4.0. Other important inputs are theparameters of fuel/assembly manufacturingclassified as the engineering parameters, which are crucial to the simulation model.The uncertainty of these parameters can propagate to the result through the simulating model and affect the accuracy of the result.

The code SUACL based on Monte Carlo sampling method wereprogrammed to dothe uncertainty analysis of cross section.The Monte Carlo sampling method and the SUACL is introduced in section 2. The perturbation model of cross section and parameters of fuel/assembly manufacturing is in section 3. All results and discussions aredescribedin section 4. Summary is in the last section.

## 2. MC Sampling Method and SUACL

### 2.1 Monte Carlo Sampling Method
The implementation of MC sampling method can often be performed as follow:[4]: define an appropriate probabilitydistribution function of inputs; generate enough analysis samples in the distribution;deliver the samples to neutron physics calculation; determine the uncertainty of desired output; determine the sensitivity of the desired output.The sample size is calculated by Wilks' formula (1) [5].

$$(1-\alpha^N) - N(1-\alpha)\alpha^{N-1} \geq \gamma \qquad (1)$$

Here, N is the minimum sampling number, $\gamma$ is co nfidence coefficient.

### 2.2 The code of SUACL
The code SUACL has been developed to perform the un certainty analysis of cross section. The procedure of unc ertainty analysis like introduced in figure 1.

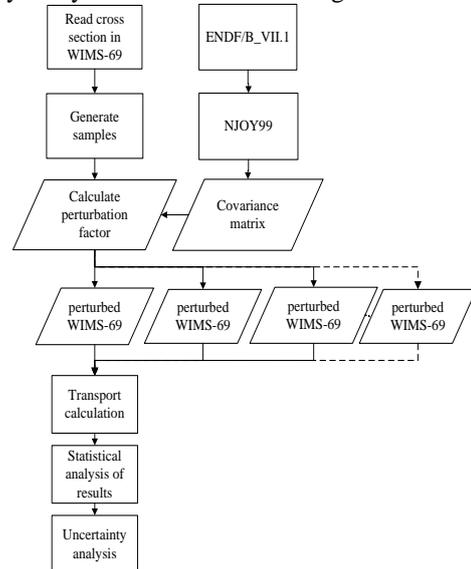

Fig1. The flowchart of SUACL

## 3 Perturbation model

### 3.1 Cross section Perturbation model
In WIMS-69, a few types of basic cross section are stored in their integral cross section like the capture cross section etc. .So before build perturbation modelof cross section, it is requiredto obtain the actual value of the cross section.It is relative to the relationship betwee n theintegral cross section and their components.The relationship is given in eq. (2)-eq. (3)

$$\sigma_a = \sigma_f + \sigma_\gamma + \sigma_\alpha + \sigma_p - \sigma_{2n} - 2\sigma_{3n} \qquad (2)$$

$$\sigma_s = \sigma_{elas} + \sigma_{inelas} + 2\sigma_{2n} + 3\sigma_{3n} \qquad (3)$$

$$\sigma_t = \sigma_f + \sigma_\gamma + \sigma_{elas} + \sigma_{inelas} + \sigma_{2n} + \sigma_{3n} \qquad (4)$$

Secondly is to obtain the perturbation factor based o n the covariance matrix. The variance and covariance o f cross section are elements kept in corresponding cova riance matrix, so its uncertainty stored in covariance ma trix.The Jacobi Rotation decomposition method is impl emented to perform the decomposition of covariance matrix. After decomposition, one diagonalizable matrix D consisted of eigenvalues and two matrices V composed by the vector of eigenvalue can be obtained. The square root of covariance matrix equal to two matrixes multiplies the square root of diagonalizable matrix $D^{\frac{1}{2}}$. All computations are shown in eq. (5)-eq. (6). The perturbation factor P is the square root of covariance matrix multiply with a vector of random number which was sampled in the standard normal distribution, and plus a unit matrix I of the same dimension with input parameter like eq. (7).

$$\Sigma = V \times D \times V^T \quad (5)$$
$$\Sigma^{\frac{1}{2}} = V \times D^{\frac{1}{2}} \times V^T \quad (6)$$
$$P = \Sigma_r^{1/2} G_{nx}(0,1) + I \quad (7)$$

The consistency rules of cross section perturbation are detailed displayed in table I to III.

Table I. The consistency rule of perturbed fission and capture cross section

| Basic cross section | Perturbation factor | Perturbed cross section |
|---|---|---|
| $\sigma_i$ (i=f/γ) | $P = 1 + \delta_i$ | $\sigma_i' = \sigma_i + \delta_i \sigma_i$ |
| | | $\sigma_a' = \delta_i \sigma_i + \sigma_a$ |
| | | $\sigma_t' = \delta_i \sigma_i + \sigma_t$ |
| | | $\sigma_{vf}' = \upsilon \delta_f \sigma_f + \sigma_{vf}$ |
| | | $I_a'(T,\sigma_0) = \frac{(1+\delta_a)\bar{\sigma}_a(T,\sigma_0')\sigma_0}{(1+\delta_a)\bar{\sigma}_a(T,\sigma_0')+\sigma_0}$ |
| | | $I_{vf}'(T,\sigma_0) = \frac{(1+\delta_{vf})\bar{\sigma}_{vf}(T,\sigma_0')\sigma_0}{(1+\delta_a)\bar{\sigma}_a(T,\sigma_0')+\sigma_0}$ |

Table II. The consistency rule of perturbed elastic scatter and inelastic scatter cross section

| Basic cross section | Perturbation factor | Perturbed cross section |
|---|---|---|
| $\sigma_i$ (i= elas / inelas) | $P = 1 + \delta_i$ | $\sigma_i' = (1+\delta_i)\sigma_i$ |
| | | $\sigma_s' = \delta_i \sigma_i + \sigma_s$ |
| | | $\sigma_t' = \delta_i \sigma_i + \sigma_t$ |
| | | $I_a'(T,\sigma_0) = \frac{(1+\delta_a)\bar{\sigma}_a(T,\sigma_0')\sigma_0}{(1+\delta_a)\bar{\sigma}_a(T,\sigma_0')+\sigma_0}$ |
| | | $I_{vf}'(T,\sigma_0) = \frac{(1+\delta_{vf})\bar{\sigma}_{vf}(T,\sigma_0')\sigma_0}{(1+\delta_a)\bar{\sigma}_a(T,\sigma_0')+\sigma_0}$ |

Table III. The consistency rule of perturbed average total fission neutron

| Basic cross section | Perturbation factor | Perturbed cross section |
|---|---|---|
| $\upsilon_{tot}$ | $P = 1 + \delta_\upsilon$ | $\sigma_{vf} = \upsilon' \sigma_f$ |
| | | $I_{vf}'(T,\sigma_0) = \frac{(1+\delta_{vf})\bar{\sigma}_{vf}(T,\sigma_0')\sigma_0}{(1+\delta_a)\bar{\sigma}_a(T,\sigma_0')+\sigma_0}$ |

### 3.2 Parameters of Fuel/Assembly Manufacturing Perturbation model

Two typical PWR cells have been recommended to quantify the uncertainty of parameters of fuel/assembly manufacturing. One is a traditional $UO_2$ cell called TMI-1, which is described in UAM benchmark. The other is created by cosRMC and filled with MOX fuel. The introduction of models is shown in table 4. According to the reference [6], the 3σ of these parameters are defined: (Fuel density: ±0.17 g/cm³; Fuel pellet diameter: ±0.013 mm; Gas thickness: ±0.024 mm; Clad thickness: ±0.025; 235U concentration: ±0.0024 w/o). In order to compare the results easily, the same distribution is used to analyze the uncertainty of parameters of MOX cell.

Table IV. The information of TMI-1 cell and MOX cell

| parameters | MOX-cell | TMI-1cell |
|---|---|---|
| Fuel material | $PuO_2$-$UO_2$(MOX) | $UO_2$ |
| Gap material | N/A | He(He4) |
| Clad material | 316SS | Zircaloy-4 |
| moderator | $H_2O$ | $H_2O$ |
| Fuel pellet[mm] | 9.020 | 9.391 |
| Gap thickness[mm] | 0.000 | 0.955 |
| Clad thickness[mm] | 0.380 | 0.673 |
| Unit cell pitch[mm] | 12.600 | 14.427 |

## 4 Results and Analysis

The paper has researched on the uncertainty of eigenvalue caused by the uncertainty of cross section and the parameters of fuel/assembly manufacturing respectively in framework of TMI-1-cell and MOX-cell. The reference codes are TSUNAMI-1D; UNICORN and SAINT. The covariance data ZZ-SCALE6.0/COVA-44G, which originate from the SACLE (based on ENDF/B-VII, ENDF/B-VI, and JENDL-3.1) is adopted by TSUNAMI-1D. The covariance matrix produced by NJOY99 is used for the UNICORN and SAINT [7-9]. All the statistical results of cross section are discussed in 4.1 and 4.2 section, while the results of parameters of fuel/assembly manufacturing are talked in section 4.3.

### 4.1 Result of Cross Section Uncertainty in TMI-1 cell

The uncertainty calculated by SUACL is in good accordance with the result of the reference code. As shown in figure 2 and figure 3, the variation trend of uncertainty of $K_{eff}$ calculated is similar to the SAINT results, when the covariance used is produced by NJOY99 by processing the ENDF/B - VII.1 and JENDL4.0. The uncertainty of $^{238}U$ capture cross section and the $^{235}U$ fission yield have occupied the majority of uncertainty source regardless of the covariance origin. The contributions to the uncertainty mainly depend on cross section of $^{238}U$ and $^{235}U$. Considering the enrichment of uranium in $UO_2$ fuel, more attention needs pay to the optimization of $^{238}U$ and $^{235}U$ cross section.

Obviously, the uncertainty of $K_{eff}$ obtained from SUACL differs from the results of TSUNAMI with the respect of capture and elastic scattering cross section of $^1H$. But the result is similar to the UNICORN ones. Comparing the energy spectrum of $^1H$ which obtained from the covariance library used by TSUNAMI and SUACL, except the $10^{-1} \sim 10^3$ of the $^1H$ elastic scattering energy spectrum, the others cross sections are just showed the same development trend and significant difference of numerical value. However, the covariance library used by SUACL is processed the ENDF/B-VII.1, which is also adopted by UNICORN analysis. It is corresponded to the theory-the uncertainty of cross section is determined by its covariance.

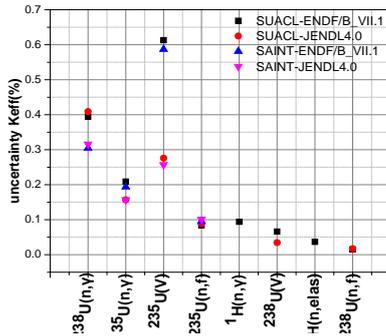 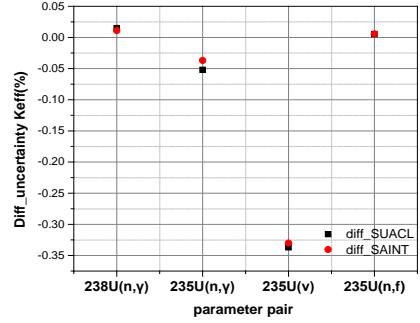

Fig.2a.The uncertainty result of $K_{eff}$ in SUACL and SAINT  Fig.2b.The different value of ENDF/B-VII.1 and JENDL4.0

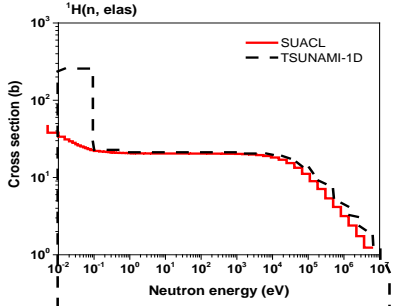 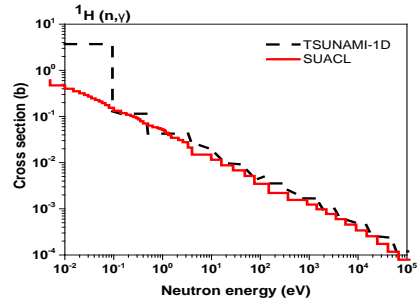

Fig.3a.The $^1$H elastic scattering energy spectrum of the covariance  Fig.3b.The $^1$H capture energy spectrum of the covariance

Table V. The uncertainty of $K_{eff}$ in TMI-1 cell (Contribution to $\frac{\Delta k}{k}$ (%))

| nuclide | Parameter pair | SUACL (ENDF/B_VII.1) | SUACL (JENDL4.0) | TSUNAMI | SAINT (ENDF/B_VII.1) | SAINT (JENDL4.0) | UNICORN |
|---|---|---|---|---|---|---|---|
| $^{238}$U | $\sigma_\gamma,\sigma_\gamma$ | 3.94E-1 | 4.09E-1 | 2.79E-1 | 3.05E-1 | 3.16E-1 | 3.77E-1 |
| $^{235}$U | $\sigma_\gamma,\sigma_\gamma$ | 2.09E-1 | 1.57E-1 | 2.11E-1 | 1.94E-1 | 1.57E-1 | 1.95E-1 |
| $^{235}$U | $v$ | 6.13E-1 | 2.76E-1 | 2.64E-1 | 5.87E-1 | 2.57E-1 | 1.97E-2 |
| $^{235}$U | $\sigma_f,\sigma_f$ | 8.40E-2 | 8.93E-2 | 7.67E-2 | 9.53E-2 | 1.01E-1 | 7.89E-2 |
| $^1$H | $\sigma_\gamma,\sigma_\gamma$ | 9.42E-2 | N/A | 1.84 E-2 | N/A | N/A | 9.56E-2 |
| $^{238}$U | $v$ | 6.61E-2 | 3.47E-2 | 7.15E-2 | N/A | N/A | 7.00E-2 |
| $^1$H | $\sigma_{elas},\sigma_{elas}$ | 3.68E-2 | N/A | 2.74E-2 | N/A | N/A | 3.79E-2 |
| $^{238}$U | $\sigma_f,\sigma_f$ | 1.52E-2 | 1.72E-2 | 1.52E-2 | N/A | N/A | 1.46E-2 |

*4.2 Result of Cross Section Uncertainty in MOX pin cell*

For MOX pin cell, filled with $PuO_2$ and $UO_2$. The contributions to the uncertainty mainly depend on cross section of $^{239}$Pu and $^{238}$U, especially the $^{238}$U capture cross section, which means the $^{238}$U play important role to determine the uncertainty of $K_{eff}$ in PWR cell.

Comparing the result with TSUNAMI, it is obvious that the total fission yield of $^{239}$Pu is less than the TSUNAMI. The reason is suspected to be similar to the capture cross section of $^1$H in TMI cell, which own to the covariance. And it is reasonable proof that the uncertainty calculated by SUACL based on the covariance originated from ENDF/B-VII.1 and JENDL4.0 is close to the SAINT. The uncertainties based on the covariance library obtained from ENDF/B-VII.1 differ from the result of JENDL4.0, which verified the relationship of covariance matrix and the uncertainty of cross section.

Table VI. The uncertainty of $K_{eff}$ in MOX cell (Contribution to $\frac{\Delta k}{k}$ (%))

| nuclide | Parameter pair | SUACL (ENDF/B_VII.1) | SUACL (JENDL4.0) | TSUNAMI | SAINT (ENDF/B_VII.1) | SAINT (JENDL4.0) |
|---|---|---|---|---|---|---|
| $^{239}$Pu | $\sigma_\gamma,\sigma_\gamma$ | 2.11E-1 | 1.88E-1 | 2.04E-1 | 1.94E-1 | 1.93E-01 |
| $^{239}$Pu | $\sigma_f,\sigma_f$ | 1.82E-1 | 2.17E-1 | 1.92E-1 | 2.38E-1 | 2.40E-01 |
| $^{239}$Pu | $v$ | 1.08E-1 | 5.20E-2 | 6.51E-1 | 1.12E-1 | 5.74E-02 |
| $^{240}$Pu | $\sigma_\gamma,\sigma_\gamma$ | 6.29E-2 | N/A | 8.84E-2 | 1.59E-1 | 5.6239E-1 |
| $^{242}$Pu | $\sigma_\gamma,\sigma_\gamma$ | 1.10E-2 | N/A | 1.19E-2 | 1.78E-1 | 5.9117E-2 |
| $^{238}$U | $\sigma_\gamma,\sigma_\gamma$ | 2.84E-1 | 2.99E-1 | 2.15E-1 | 2.16E-1 | 2.2838E-1 |
| $^{238}$U | $\sigma_f,\sigma_f$ | 1.74E-2 | 1.98E-2 | 1.80E-2 | N/A | N/A |
| $^{238}$U | $v$ | 6.79E-2 | 3.57E-2 | 7.68E-2 | 9.62E-2 | 4.8748E-2 |
| $^{235}$U | $\sigma_\gamma, \sigma_\gamma$ | 5.90E-2 | 3.99E-2 | 6.09E-2 | N/A | N/A |
| $^{235}$U | $\sigma_f, \sigma_f$ | 2.03E-2 | 3.34E-2 | 2.47E-2 | N/A | N/A |
| $^{1}$H | $\sigma_{elas}, \sigma_{elas}$ | 3.04E-2 | N/A | 3.83E-2 | N/A | N/A |

*4.3 Result of uncertainty of parameters of fuel/assembly manufacturing*

The parameters of fuel/assembly manufacturinguncertaintyfor TMI-1 and MOX cell are comparable, and their numerical value is in the same magnitudeas the uncertainty caused by some critical cross section like fission.The $^{235}$U concentration uncertainty is the most important uncertainty source to TMI-1cell. Furthermore, the cell is filled with UO$_2$and the majority of fission reaction occurs in$^{235}$U. So the $^{235}$U concentration uncertainty plays a decisive role TMI-1cell. However, compared with TMI-1 cell, no gas in MOX cell and its whole area is less than the TMI-1, so the impact of clad is greater in it. The fuel density uncertainty has great effect on the eigenvalue uncertainty in typical PWR cell.

Table VII.The uncertainty of K$_{eff}$caused by parameter in TMI-1cell and MOX cell

| Parameter of geometry and fuel | Contribution to $\frac{\Delta k}{k}$ (%), TMI_1 cell | Contribution to $\frac{\Delta k}{k}$ (%), MOX cell |
|---|---|---|
| Fuel density | 4.65E-2 | 3.8E-2 |
| Fuel pellet diameter | 1.78E-2 | 2.07E-2 |
| Gas thickness | 3.24E-2 | N/A |
| Clad thickness | 4.77E-2 | 1.37E-1 |
| $^{235}$U concentration | 1.60E-1 | N/A |

## 5 Conclusions

The uncertainty of various cross sections and manufacturing parameters of fuel/assembly have been analyzed by implementing the Monte Carlo method. Two typical PWR models were constructed to verify the SUACL based on different covariance library.

According to the result of cross section,all results of SUACL werefound in accordance with the resultsof reference codes.$^{238}$U and $^{235}$U play an important role in determining the uncertainty of K$_{eff}$in TMI-1 cell.The uncertainty of K$_{eff}$in MOX ismainly affected by $^{239}$Pu and $^{238}$U. It is obvious that the uncertainty mostly depends on the covariance library and insensitivities to cross section library. The uncertainties based on the covariance library obtained from ENDF/B‐VII.1 differ from the result of JENDL4.0, which verified the relationship of covariance matrix and the uncertainty of cross section. And the parameters of fuel/assembly manufacturing uncertainty are comparable to uncertainty of some cross section, especially the 235U concentration, clad thickness. The uncertainty analysis of these parameters is of great significance to evaluate the parameter of actual cell and help to improve the modelsimulated. More attentions need paid to improve the accuracyof the parameters analysis talked in the paper.

**Acknowledgements**

The work was supported by National Natural Science Foundation of China (No. 11390383) and the Fundamental Research Funds for the Central Universities (No. 2015ZZD12), we would like to thankHafiz Muhammad Waqar Sharif.